# Mass Formulas Derived by Symmetry Breaking and Prediction of Masses on Heavy Flavor Hadrons

Yi-Fang Chang
*Department of Physics, Yunnan University, Kunming 650091, China*
(E-mail: yifangchang1030@hotmail.com)

**Abstract:** The base is the Lagrangian of symmetry and its dynamical breaking or Higgs breaking. When the soliton-like solutions of the scalar field equations are substituted into the spinor field equations, in the approximation of non-relativity we derive the Morse-type potential, whose energy spectrum is the GMO mass formula and its modified accurate mass formula $M = M_0 + AS + B[I(I+1) - S^2/2]$. According to the symmetry of s-c quarks, the heavy flavor hadrons which made of u,d and c quarks may be classified by SU(3) octet and decuplet. Then some simple mass formulas are obtained, from this we predict some masses of unknown hadrons, for example, m($\Xi_{cc}$)=3715 or 3673MeV, and m($\Omega_{cc}^+$)=3950.8 or 3908.2MeV, etc.
**Key words:** hadron, mass formula, symmetry breaking, heavy flavor
**PACS** 12.70. +q, 11.30. Hv

## 1. Introduction

It is very successful that the SU(3) symmetry and its broken are applied to the classification of the ground hadrons, which made of u, d and s quarks, and to the mass spectrum, etc. It is also well known that those hadrons agree with the GMO mass formula

$$M = M_0 + AS + B[I(I+1) - \frac{S^2}{4}]. \qquad (1)$$

Then there is a simple mass formula [1]

$$M = M_0 + \alpha Y + \beta C + \gamma[\frac{Y^2}{4} - I(I+1)] + \delta J(J+1). \qquad (2)$$

for the SU(8) supermultiplets. The hadron spectrum of the quenched QCD was calculated [2]. The mass spectrum of supersymmetric generalization of QCD was studied [3]. The algebraic methods of hadron spectrum were reviewed [4]. The baryon mass spectrum in a SU(3) hidden gauge symmetry was calculated [5]. The spectrum of baryons with two heavy quarks and the mass spectrum of three-generation models were analysed [6,7]. Recently, Anderson obtained the heavy quark mass scale, then calculated and predicted some masses of light and heavy quark hadrons [8]. Takenaka and Sakai described the mass formula for the resonances by the micrononcausal Euclidean wave functions [9]. Lichtenberg, et al., obtained the formulas of masses of ground-state hadrons, most of which contain heavy-flavor quarks, by the regularities in hadron interactions [10]. Glozman and Riska described the spectrum of the charm hyperons in a chiral quark model [11]. Forcrand, et al., reported the strange and charmed hadron spectroscopy on a lattice theory [12]. Cui solved the BS equation for the hybrid mesons with heavy quarks under instantaneous approximation, and obtained the spectrum of hybrid mesons [13]. Genovese, et al., obtained isospin mass of baryons in potential models [14].

In this paper, based on the symmetry and its breaking, the mass formulas are derived, from which the masses of some heavy flavor hadrons are predicted.

## 2. Breaking Theory on Symmetry and Mass Formulas

The usual Lagrangian of the gauge field of SU(3) symmetrical strong interaction is [15,16]

$$L = -\frac{1}{4}F_{\mu\nu}F^{\mu\nu} - \overline{\psi}\gamma^\mu(\partial_\mu + ig\gamma^5 A_\mu)\psi - \overline{\psi}m\psi. \qquad (3)$$

When the vector field $A_\mu$ has mass $\mu$, the phenomenological Lagrangian of dynamical breaking is [16]

$$L = -\frac{1}{4}F_{\mu\nu}F^{\mu\nu} + \frac{1}{2}\mu^2 A_\mu A^\mu - \overline{\psi}\gamma^\mu(\partial_\mu + ig\gamma^5 A_\mu)\psi - \frac{1}{2}\partial_\mu\varphi\partial^\mu\varphi$$
$$-m\overline{\psi}\exp(\frac{2g\gamma^5}{\mu}\varphi) + \mu A^\mu \partial_\mu\varphi. \quad (4)$$

Hence the equations of motion are

$$\gamma^\mu(\partial_\mu + ig\gamma^5 A_\mu)\psi + m\exp(\frac{2g\gamma^5}{\mu}\varphi)\psi = 0. \quad (5)$$

$$\Box\varphi = \mu\partial_\mu A^\mu + \frac{2g\gamma^5}{\mu}m\exp(\frac{2g\gamma^5}{\mu}\varphi)\overline{\psi}\psi. \quad (6)$$

$$\partial_\nu F^{\mu\nu} + gf^{ab}A_{a\nu}F_b^{\mu\nu} = -\mu^2 A^\mu - \mu\partial^\mu\varphi + ig\overline{\psi}\gamma^\mu\gamma^5\psi. \quad (7)$$

When $A^\mu = 0$, Eq.(6) becomes

$$\Box\varphi = \frac{2g\gamma^5}{\mu}m\overline{\psi}\psi\exp(\frac{2g\gamma^5}{\mu}\varphi) = ae^{b\varphi}. \quad (8)$$

Let the integral constant is positive (e.g.,1/2), we derive a particular solution

$$\varphi = \frac{1}{b}\ln\frac{b}{2a}\frac{4D}{(1-D)^2}, D = \exp(\pm b\frac{r-ut}{\sqrt{1-u^2}} + C'), \quad (9)$$

which is analogous with the soliton solution. We give up the meaningless $D=exp(+b....)$, then $D<1, e^{b\varphi} = 2bD/a(1-D)^2$. In the approximation of non-relativity, $\sqrt{1-u^2} \approx 1, D = \exp[-b(r-r_0)], D/(1-D)^2 \approx D$, so

$$e^{b\varphi} = \frac{2b}{a}D = C\exp[-b(r-r_0)]. \quad (10)$$

Eq.(5) is left multiplication by $[\gamma^\mu\partial_\mu + m\exp(2g\gamma^5\varphi/\mu) - ig\gamma^\mu\gamma^5 A_\mu]$, we obtain

$$[\Box + m^2\exp(\frac{4g\gamma^5}{\mu}\varphi) + 2m\exp(\frac{2g\gamma^5}{\mu}\varphi)\gamma^\mu\partial_\mu]\psi - g^2 A_\mu^2\psi + 2ig\gamma^5 A_\mu\partial_\mu\psi = 0. \quad (11)$$

Because the hadrons possess the SU(3) and SU(6) symmetries, the internal structure of these particles is of low velocity, then their momentums can be neglected in non-relativity. When $A_\mu = 0$, the total energy $E = E'+m$, Eq.(11) becomes

$$\frac{d^3 S}{dr^2} + \{-\frac{K(K+1)}{r^2} + 2mE' + m^2[1-\exp(\frac{2g\gamma^5}{\mu}\varphi)]^2\}S = 0. \quad (12)$$

The approximate solution (10) of the scalar field $\varphi$ equation is replaced into Eq.(12), then the potential is the Morse-type function [17] $U(r) = -m[1-Ce^{-b(r-r_0)}]^2/2$, the equation is

$$\frac{d^2 S}{dr^2} + [-\frac{K(K+1)}{r^2} + 2m(E'-U)]S = 0. \quad (13)$$

Its energy level is

$$E'_{k,n} = E'_{0,0} + An + BK(K+1) + Cn^2 - DK^2(K+1)^2 + EnK(K+1). \quad (14)$$

Where let $\mu = i\mu_0$, so $E_{0,0} = -(g/\mu_0) + (g^2/2m\mu_0^2)$, $A = -(2g/\mu_0) + (2g^2/m\mu_0^2)$, $B = 1/(2mr_0^2)$, $C = 2g^2/m\mu_0^2$, etc.

The Lagrangian of Higgs breaking is [18]

$$L = -\frac{1}{4}F_{\mu\nu}F^{\mu\nu} + \frac{1}{2}\mu^2 A_\mu A^\mu - \overline{\psi}\gamma^\mu(\partial_\mu + ig\gamma^5 A_\mu)\psi - m\overline{\psi}\psi - \frac{1}{2}\partial_\mu\varphi\partial^\mu\varphi -$$
$$\frac{1}{2}e_0^2 A_\mu A^\mu \varphi^2 + \frac{1}{2}m_0^2\varphi^2 - \frac{1}{4}f^2\varphi^4. \quad (15)$$

Assume that the interaction between $\varphi$ and $\psi$ fields is $m_0^2 = 2a\bar{\psi}\psi$, which is analogous with result of dynamical breaking. Hence the corresponding equations of motion are

$$\gamma^\mu(\partial_\mu + ig\gamma^5 A_\mu)\psi + m\psi - a\varphi^2\psi = 0. \tag{16}$$

$$\Box\varphi + (m_0^2 - e_0^2 A_\mu A^\mu)\varphi - f^2\varphi^3 = 0. \tag{17}$$

$$\partial_\nu F^{\mu\nu} + gf^{ab}A_{a\nu}F_b^{\mu\nu} = (-\mu^2 + e_0^2\varphi^2)A^\mu + ig\bar{\psi}\gamma^\mu\gamma^5\psi. \tag{18}$$

If $f$ has a relation with $\psi$, Eq.(16) will add a $\varphi^4$ term, etc. When $A_\mu = 0$, Eq.(15) reverts to the Goldstone Lagrangian [19], Eq.(17) is

$$\Box\varphi = -m_0^2\varphi + f^2\varphi^3. \tag{19}$$

Let the integral constant $C = m_0^4/4f^2$ and $|\varphi| < m_0/f$, its soliton solution is

$$\varphi = \frac{m_0}{f}\frac{F-1}{F+1}, F = \exp(\pm\sqrt{2}m_0\frac{r-ut}{\sqrt{1-u^2}} + C'). \tag{20}$$

It is just a kink of the $\varphi^4$ equation. When the solution $F = \exp(+\sqrt{2}m_0....)$ is neglected,

$$F < 1, \varphi \approx (m_0/f)(-1 + 2F - 2F^2).$$

Eq.(16) is left multiplication by $(\gamma^\mu\partial_\mu + m - a\varphi^2 - ig\gamma^\mu\gamma^5 A_\mu)$, so we obtain

$$[\Box + m^2 + 2m\gamma^\mu\partial_\mu - 2a\varphi^2(m + \gamma^\mu\partial_\mu) + a^2\varphi^4]\psi - g^2 A_\mu^2\psi + 2ig\gamma^5 A_\mu\partial_\mu\psi = 0. \tag{21}$$

When $A_\mu = 0, E = m + E'$, Eq.(21) becomes to Eq.(13) for non-relativity. The potential is

$$U(r) = -(a^2/2m)\varphi^4 = -(a^2/2m)(1 - 8F + 32F^2), \tag{22}$$

where $F = \exp[-\sqrt{2}m_0(r-ut) + C_0] = \exp[-\sqrt{2}m_0(r-r_0)]$. This is also the Morse-type function. Hence the energy level is still the formula (14). It is the same with the mass formula

$$M = (m + E'_{0,0}) + AS + BI(I+1) + CS^2, \tag{23}$$

and whose two order form. In this case,

$$M_0 = m + E'_{0,0}, E'_{0,0} = (m_0^2/m)[(a^2 m_0^2/2f^4) + (iam_0/f^2) - (1/4)],$$

$$A = (m_0^2/m)[(2iam_0/f^2) - 1], B = \hbar^2 c^2/2mr_0^2, C = -m_0^2/m.$$

For the $J^P = 1^+/2$ baryon octet, let

$$a/f^2 = 0.03178i, m=1416.447\text{MeV}, r_0 = 0.598\times 10^{-13} cm, m_0 = 156.0408\text{MeV}.$$

Such Eq.(23) agrees completely with the experimental data. If C=-B/2, i.e., $m_0 = \hbar c/2r_0 = (98.66429/r_0)$ Mevfm, Eq.(23) will be the mass formula [20,21]

$$M = M_0 + AS + B[I(I+1) - \frac{S^2}{2}]. \tag{24}$$

Let $M_0 = 908, A = -228,$ and B=40MeV, so

$$m(N) = 938, m(\Lambda) = 1116, m(\Sigma) = 1196, m(\Xi) = 1314 \text{ MeV}.$$

For the $J^{pc} = 0^{-+}$ meson octet, let A=0, $M_0$=549.4MeV and B=--207.22MeV, so

$$m(\pi^0) = 134.96, m(K^0) = 497.6, m(\eta) = 549.4 \text{ MeV}.$$

The neutral mesons agree completely within the range of error, and M=m. For the $J^P = 3^+/2$ baryon decuplet I=1+(B+S)/2 always holds, so the formula (24) just derives a more accurate formula

$$M = M_0 + aS + bS^2 = M'_0 + AY + CY^2. \tag{25}$$

Let $M'_0 = 1385.3, A = -150.2$ and C=-3.3MeV, so

$$m(\Delta) = 1231.8, m(\Sigma) = 1385.3, m(\Xi) = 1532.2, m(\Omega) = 1672.5 \text{ MeV}.$$

The mass relations

$$2[m(N) + m(\Xi)] = 3m(\Lambda) + m(\Sigma), \tag{26}$$

$$4[m(N) + m(\Xi)] = 7m(\Lambda) + m(\Sigma), \tag{27}$$

$$8m(K) = 7m(\eta) + m(\pi). \tag{28}$$

They correspond to Eqs.(1) and (24), respectively.

**3. Heavy Flavor Hadrons and Their Masses**

In the standard model quarks are the three generations (u,d) (c,s) and (t,b), whose properties exhibit a better symmetry. Some hadrons including heavy flavor c, b and t quarks have been found, and they are consistent with the SU(N) multiplets. Based on the symmetry of s and c quarks in the same generation, we can suppose that the hadrons, which made of u, d and c quarks, are also the SU(3) symmetry. It is a subgroup of SU(4) of u, d, s and c quarks. Such the eight $J^P = 1^+/2$ baryons:

p=uud, n=udd(I=1/2); $\Lambda_c^+ = udc(I=0); \Sigma_c^{++} = uuc, \Sigma_c^+ = udc, \Sigma_c^0 = ddc(I=1);$

and $\Xi_{cc}^{++} = ucc, \Xi_{cc}^+ = dcc(I=1/2)$

form an octet too. Since m(N)=939MeV, $m(\Lambda_c^+) = 2285$ MeV, $m(\Sigma_c) = 2454$MeV, and

$$\frac{m(\Sigma) - m(\Lambda)}{m(\Sigma)} = 0.0645 \approx \frac{m(\Sigma_c) - m(\Lambda_c)}{m(\Sigma_c)} = 0.0681, \tag{29}$$

and $\Sigma_c \to \Lambda_c^+ \pi$ is similar to $\Sigma^0 \to \Lambda \pi$, so we assume that these masses of the octet obey a corresponding mass formula

$$M = M_0 + AC + B[I(I+1) - \frac{C^2}{4}], \tag{30}$$

or

$$M = M_0 + AC + B[I(I+1) - \frac{C^2}{2}]. \tag{31}$$

From the two corresponding mass relations we may predict $m(\Xi_{cc}) = 3715$ or 3673MeV.

For the $J^P = 1^+/2$ decuplet baryons, m($\Omega_c^0$ =ssc)=2698MeV, m($\Xi_c^0 = dsc, \Xi_c^+ = usc$)= 2576MeV, $m(\Sigma_c) = 2454$MeV, these masses are an equal-spacing rule, so the corresponding mass($\Delta$) should be 2332MeV.

For the $J^P = 3^+/2$ baryons, $\Delta^{++}, \Delta^+, \Delta^0, \Delta^-$ (I=3/2); $\Sigma_c^{++}, \Sigma_c^+, \Sigma_c^0$ (I=1); $\Xi_{cc}^{++}, \Xi_{cc}^+$ (I=1/2) and $\Omega_{ccc}^{++}$=ccc (I=0) form a decuplet too. It is the first series of heavy flavor baryons. Their masses are possibly an equal-spacing rule, i.e.

$$M = M_0 + aC. \tag{32}$$

The $J^P = 0^-$ octet of heavy flavor mesons are $\pi^{+,-}, \pi^0 (I=1); D^+ = c\bar{d}, D^0 = c\bar{u}$ (I=1/2) and their antiparticles; $\eta'_c = a(u\bar{u} + d\bar{d}) + b(c\bar{c})$. If their mass relation is 4m(D)= $m(\pi) + 3m(\eta'_c)$ or 8m(D)= $m(\pi) + 7m(\eta'_c)$, so $m(\eta'_c) = 2444$ or 2114MeV since m($\pi$)=137, m(D) =1867MeV. For the $J^P = 1^-$ octet, $m(\rho) = 770, m(D^+) = 2010$ MeV, so $m(\eta'_c) = 2423$ or 2187MeV.

These octets and decuplet are a certain cross section of the diagrams of the SU(4) multiplets, respectively. probably, the masses of the triplet $\Sigma^+, \Sigma^0, \Sigma^-$ (I=1), the doublet $\Xi_c^+ = usc, \Xi_c^0 = dsc$ (I=1/2) and the singlet $\Omega_{cc}^+ = scc$ (I=0) are approximately an equal-spacing rule, it is the second series of series of heavy flavor baryons. Then the masses of $\Omega^-, \Omega_c^0 = ssc, \Omega_{cc}^+$ and $\Omega_{ccc}^{++}$ should be equal-spacing too. It is the third series of heavy flavor baryons. Such based on the known masses of $3^+/2$ baryons including c quark [22], other masses of baryons will be able to be estimated. Because $\Sigma_c$(uuc,2518)- $\Delta$(uuu,1232)=1286≌c-u, so m($\Xi_{cc}$)=3804 and m($\Omega_{ccc}^{++}$)=5090. From the third series, we obtain m($\Omega^-$)=1675.4,

m($\Omega_c^0$)=2811.6, m($\Omega_{cc}^+$)=3950.8 and m($\Omega_{ccc}^{++}$)=5090. But, from the second series the known m($\Sigma$)= 1385 and. m($\Xi_c$)=2646.6, so c-u≅1261.6 and m($\Omega_{cc}^+$)=3908.2. Both m($\Omega_{cc}^+$) are different. Of course, these baryons obey probably the more accurate mass formulas

$$M = M_0 + aC + bC^2, M = M_0 + a'I + b'I^2,  \quad (33)$$

which correspond to Eqs.(24), (25) and (31).

Some similar multiplets and mass formulas exist possibly in baryons and mesons including b or t quarks. For instance, both mass spectra of $\psi = c\bar{c}$ and $\gamma = b\bar{b}$ are similar; as in the neutral kaon system, $D^0 - \overline{D^0}$ and $B^0 - \overline{B^0}$ mixings should exist. Therefore, the above method will be able to be developed.

**Appendix**

Based on the known theories [23,24], the mass operator of the first broken SU(3) symmetry is [23]

$$O_M = H_{VSI} + H_{MSI} = M_0 + AU_3 = M + AY + BI(I+1) + CY^2. \quad (34)$$

For the famous mass relation (26), C=-B/4 is obtained. For a new mass relation (27) which agrees better, we may obtain C=-B/2 and Eq.(24). In Ref.24, the mass of the same multiplet "can be split into irreducible parts according to the Clebsch-Gordan series

$$8 \otimes 8 = 1 \oplus 8' \oplus 8'' \oplus 10 \oplus \overline{10} \oplus 27.$$

Neutrality means that" 10 and $\overline{10}$ are absent, then

$$m_N = (m_1 - 2m'_8 + m''_8 - 3m_{27}), m_\Lambda = (m_1 - m'_8 - m''_8 + 9m_{27}), \quad (35)$$
$$m_\Sigma = (m_1 + m'_8 + m''_8 + m_{27}), m_\Xi = (m_1 + m'_8 - 2m''_8 - 3m_{27}). \quad (36)$$

"Suppose that the mass splitting is generated in a more fundamental Lagrangian by a quark-mass term of the type $\bar{\psi}(a + b\lambda_8)\psi$ carrying only singlet and octet representations and that this property is (miraculously) preserved by the interactions. This would mean that $m_{27}$ vanishes,"± leading to Eq.(26). If we let $m_{27} = (m'_8 + m''_8)/44 \neq 0$, Eq.(27) will be derived. Furthermore, the primary cause is that Eqs.(24) and (27)(28) agree better with the data, and baryon and meson can be unified by M=m.

**References**
1.J.W.Maffat, Phys.Rev.D12(1975)288.
2.R.D.Loft and T.A.DeGrand, Phys.Rev.D39(1989)2678.
3.L.Leroy, Z.Phys.C43(1989)159.
4.F.Iachello, Nucl.Phys.A497(1989)23c.
5.B.Kleihaus, J.Kunz and T.Reitz, Phys.Rev. D42(1990)1661.
6.M.J.Savage and M.B.Wise, Phys.Lett.B248(1990)177.
7.F.del Aguila et al, Nucl.Phys.B251(1991).
8.J.T.Anderson, Fortschr.Phys.42(1994)195;487.
9.A.Takenaka and I.Sakai, Nuovo Cimento,109A(1996)301.
10.D.B.Lichtenberg,et al., Phys.Rev.,D53(1996)6678.
11.L.Ya.Glozman and D.O.Riska, Nucl.Phys.,A603(1996)326.
12.Ph.de Forcrand,et al., Nucl.Phys.,B460(1996)413.
13.J.Y.Cui, Int.J.Mod.Phys., A14(1999)2273.
14.M.Genovese, J.M.Richard, B.S.Brac, et al., Phys.Rev., D59(1999)014012.
15.S.Weinberg, Phys.Rev.D8(1973)605.
16.R.Jackiw and K.Johnson, Phys.Rev.D8(1973)2386.
17.P.M.Morse, Phys.Rev.34(1929)57.
18.J.Bernstein, Rev.Mod.Phys.46(1974)7.
19.J.Goldstone, Nuov.Cim.19(1961)154.
20.Yi-Fang Chang, Hadronic J., 7(1984)1118.
21.Yi-Fang Chang, New Research of Particle Physics and Relativity. Yunnan Science and Technology Press. (1989)p1-183; Phys.Abst., 93(1990)No1371.

22. W-M Yao, C.Amsler, D.Asner, et al. Particle Data Group. J.Phys. G33(2006)1.
23. B.I.Feld, Models of Elementary Particles. Blaisdell Publishing Company(1969).
24. C.Itzykson and C.Zuber, Quantum Field Theory. McGraw-Hill(1980).